\documentclass[12pt]{iopart}
\usepackage{graphicx}

\begin{document}

\title[Role of time-varying magnetic field on QGP equation of state]{Role of time-varying magnetic field on QGP equation of state}

\author{Yogesh Kumar$^{1}$, Poonam Jain$^{2}$, Pargin Bangotra$^{3}$, Vinod Kumar$^{4}$, D. V. Singh$^{5}$ \& S. K. Rajouria$^{6}$ }

\address{$^{1}$Department of Physics, Hansraj College, University of Delhi, Malka Ganj, North Campus, New Delhi-110007, India \\
$^{2}$Department of Physics, Sri Aurobindo College, University of Delhi, Malviya Nagar, New Delhi-110017, India\\
$^{3}$Department of Physics, Netaji Subhas University of Technology, Dwarka, Delhi, India\\ 
$^{4}$ Department of Physics, University of Lucknow-226007, Lucknow, U.P., India
\\
$^{5}$ Department of Physics, GLA University, Mathura, U.P., India
\\
$^{6}$ Department of Physics, Zakir Hussain Delhi College, University of Delhi, Jawaharlal Nehru Marg, New Delhi-110002, India
}
\ead{$^{1}$yogesh.du81@gmail.com}
\vspace{10pt}

\begin{abstract}
 
The phase diagram of quantum chromodynamics (QCD) and its associated thermodynamic properties of quark gluon plasma (QGP) are studied in the presence of time dependent magnetic field. The study plays a pivotal role in the field of cosmology, astrophysics, and heavy ion collisions. In order to explore the structure of quark gluon plasma to deal with the dynamics of quarks and gluons, we investigate the equation of state (EoS) not only in the environment of static magnetic field but also in the presence of time-varying magnetic fields. So, for determining the equation of state of QGP at non zero magnetic fields, we revisited our earlier model where the effect of time varying magnetic field was not taken into consideration. Using the phenomenological model, some appealing features are noticed depending upon the three different scales; effective mass of quark, temperature, and time independent as well as time-dependent magnetic field. Earlier the effective mass of quark was incorporated in our calculations and in the current work, it is modified for static and time-varying magnetic fields. Thermodynamic observables including pressure, energy density, entropy, etc. are calculated for a wide range of temperature and time-dependent as well as time-independent magnetic fields. Finally, we claim that the EoS are highly affected in the presence of a magnetic field. Our results are notable compared to other approaches and found to be advantageous for the measurement of QGP equation of state. These crucial findings with and without time-varying magnetic field could have phenomenological implications in various sectors of high energy physics. 
\end{abstract}

%
%
%
%

\noindent{\it Keywords\/}: equation of state; quark gluon plasma; quantum chromodynamics
\section{Introduction}

The extensive research in the field of high energy physics unveiled the phase diagram of quantum chromodynamics (QCD) which involves hadronic phases (HP) as well as quark gluon plasma (QGP) phase \cite{z4,z6,z8,z9,z10,z19,z26,z30,z36,z43}. This area captivated some light on revealing the process of phase transition at high temperatures and densities in heavy ion collision experiments~\cite{sto,lacey}. QCD is a particular theory related to the strong interactions of quarks and gluons. It is observed that at the critical temperature $T_c\approx 170$ MeV, the hadronic matter transforms into a new phase where quarks and gluons are almost free called quark gluon plasma (QGP) \cite{harris,brau,gia,kar}. The accelerator (RHIC and LHC) facilities around the globe provide an opportunity to look deep into the matter where several unresolved mysteries for the formation and evolution of the universe after the big bang may resolve. The upcoming facility (FAIR) at GSI is also working in the same direction in which physicists are trying to gain some useful insights into the basic structure of matter and investigate the evolution of the universe. 
\par 
One of the best measurements of these accelerators are to deal with the properties of QGP in which the nature of phase transitions, critical temperature, and thermodynamic properties are treated as a peculiar study. Besides these, equation of state (EoS) are one such characteristic of QGP that encodes the thermodynamic properties of the system. Even after rigorous work done on theory and experiment over the couple of past decades, our understanding of QGP signals and its related properties are still not well understood \cite{shury,satz}. 
\par 
It has been claimed that transitions between quark-hadronic phases may have transpired during the initial moments of the early universe, as the system underwent expansion and cooled within a few femtoseconds following the Big Bang. A similar scenario is anticipated to arise during relativistic heavy ion collisions at facilities like RHIC and LHC, giving rise to a miniature version of the cosmic conditions. Exploring the interplay of thermodynamic properties during these phase transitions is intriguing. The thermodynamic properties, essential components of the EoS, showcase how dynamic parameters like chemical potential and temperature contribute to the transformative process of the system into distinct phases.
\par 
Another important physical quantity noticed by some physicists at RHIC and LHC is the magnetic field of the order of $\sim 10^{15}$ Tesla that exists from the microscopic level to the macroscopic level. A surprising magnetic phenomenon emerges in the transverse direction of the reaction plane when heavy ion beams collide at relativistic speeds \cite{a3,a5,kharzeev}. The QCD phase structure is influenced not only by temperature and chemical potential but also by another crucial parameter; the magnetic field. The equation of state exhibits intriguing behavior when probing the complexities of the quark-gluon plasma  system in the presence of such magnetic fields. Since a large amount of magnetic field produced due to the spectators, its strength is surprisingly approach to the mass of hadrons. So, the strength of the magnetic field created at RHIC and LHC is a few times of hadron mass \cite{a7,a8}. Numerous physicists have postulated its creation during the evolution of the early universe \cite{grasso}. Astrophysicists similarly posit the existence of strong magnetic fields in celestial bodies such as neutron stars, boson stars, strange quark matter, and beyond \cite{s2,s5,s6,s8,s10,s14}. It is pointed out that an enormous amount of magnetic field $\sim$ $10^{19}-10^{20}$ $G$ might be generated in heavy-ion collisions \cite{f2,f3,f4,f5} that may influence the various properties of QGP matter drastically. It is also very difficult to deal with such an enormous magnetic field generated in the collisions of heavy ions  due to its existence for a very little span of time \cite{f9}. Some other more interesting features have been explored in the case of dealing with chiral magnetic effect \cite{f10,f13,f15} and the magnetic catalysis effects \cite{f20,f21,f22}.
\par 
Results from Lattice QCD investigations \cite{bali3} have scrutinized the impact of external magnetic fields on the EoS of QGP. The study involved the evaluation of thermodynamic variables such as pressure, entropy, and energy density. Authors in related works \cite{bali1, agasian, endrodi, ayala1, ayala3} observed a reduction in the transition temperature under the influence of a magnetic field. Similarly, employing low-energy effective theories, the robust magnetic field is demonstrated to exert an influence on the phase diagram of hadronic matter \cite{andersen, fraga}. It is now clear that the deep understanding of EoS based on chemical potential, temperature, and magnetic field are the main requirement to investigate the phase diagram of QCD. On the other hand, lattice simulation data is still found limited, i.e., valid at zero chemical potential and magnetic field \cite{b1,b2,b3}. Due to the sign problem, simulation data with finite chemical potential is difficult to obtained, whereas Monte-Carlo simulation works appreciably at finite magnetic fields, yet the results from lattice data with finite magnetic fields are not very much clear in order to produce the EoS. 
\par 
Until now, the authors have employed a widely recognized and straightforward model known as the bag model to examine and elucidate the characteristics of QGP \cite{shur, cley}. These studies inspired us to investigate EoS in the relevant range of magnetic fields and to see the behaviour of time-varying magnetic field on QGP EoS. Here, our main aim is to determine the QGP EoS with a wide range of temperatures incorporating the effective quark mass at zero chemical potential and then modify our calculations in the presence of time-independent and time-varying magnetic field. To this end, we develop a theoretical model to study the thermodynamic variables like pressure, the energy density, the entropy, etc. with the finite values of magnetic field and compare these results with the vanishing magnetic field. In addition to this, we also check the validity of time-varying magnetic field on QGP EoS. 
\par 
Above all, we first started the work in the zero limit of chemical potential which is examined as a cogent approximation for the study of hot QGP in relativistic heavy ion collisions experiments at BNL (RHIC), CERN (LHC) and upcoming FAIR (GSI). Till date, Lattice QCD results are valid at zero value of chemical potential for the study of QGP EoS. In this manuscript, an effective mass of the quark is used in place of the thermal quark mass as a valid approximation of the quasiparticle model assuming zero chemical potential at RHIC and LHC. This model has been consistently applied in a self-contained manner. It's crucial to emphasize that addressing the various divergences inherent in the computation of free energies for quarks and gluons is achieved through the inclusion of appropriate counter terms. The current work predicated upon our prior research, illustrating the significance of the quark mass's temperature dependence in revealing intriguing behaviors within the QGP EoS during heavy-ion collisions. The calculations are further refined by incorporating magnetic fields and adjusting the effective quark mass accordingly. This underscores the role of theoretical models as indispensable tools for exploring the QGP EoS, particularly in the intricate context of magnetic fields.
\par 
The current studies are found to exhibit some keen features of a thermodynamic system in the presence of a magnetic field with the involvement of multiple scales, i.e, effective quark mass, temperature and constant and time-varying magnetic field. One can work with two limiting cases: the strong magnetic field limit or low temperature and the weak magnetic field limit or high temperature. Inclusion of a magnetic field in the medium requires a suitable modification of the present theoretical tools to investigate various properties of QGP. The important features of the Landau levels are the separation of the levels proportional to the magnetic field. Considering high and intense magnetic fields, a weak interaction between the charged particles only perturbs the Landau levels and then the lowest Landau level (LLL) dominance holds good in such cases. So, the LLL approximation is physically the most suitable and widely employed by several authors in strong magnetic fields \cite{q2,q4,q6,q7}, while the contributions from the highest Landau levels are difficult to estimate \cite{q8,q9,q10,q11} in such field limits. Recently, an interesting work has been done by Tabatabaee and Sadooghi who have used Wigner function formalism and evaluated the thermodynamic quantities in an expanding magnetized plasma incorporating the quantum effects arising from Landau quantization \cite{q12}.
\par 
Another group \cite{q13} has also reported that the magnetic field pushes higher Landau levels to infinity compared to the lowest Landau level in the strong  field limit. Then thermally, quarks excite very feebly to the higher Landau levels and only the LLL, $n=0$, are populated. This shows that the system is considered confined in the LLL for such strong field cases. Moreover, the reason for this assumption can also be understood by looking at the dispersion relation in the presence of an external magnetic field along z-direction. Finally, in the current work, we deal with the LLL approximation along with the associated scales; $eB$, $T$ and $m_{eff}$.
\par
Thus, our paper is organised in three different sections. It is as follows: in Section $II$, we briefly explain the importance of our model. In Section $III$, we show the results of free energy that can help in order to produce the EoS of QGP. After this, the results are presented in Section $IV$ and at last, we conclude our work in Section $V$.
\par 
\section{\bf Description of a theoretical model}
QGP's formartion in relativistic heavy ion collisions has now become the main attraction of high energy physicists. Among several properties of QGP, the equation of state of QGP is widely explored and the most popular study as the best signature of QGP. Numerous researchers posit that the EoS serves as an indirect yet valuable signature for elucidating the formation and evolution of QGP. In this quest, various theoretical models align closely with calculations from both perturbative and non-perturbative QCD, making them thermodynamically sound for unraveling the intricate structure of QGP \cite{lev, sch}. 
\par 
Significant strides have been made in the exploration of QGP through the quasiparticle approach \cite{bb}. Authors have used this model which is well suited just above and around the critical temperature where hadronic matter is supposed to have a phase transition and after a valid transition, it forms a new state of matter. This new state of matter is having free quarks and gluons. Initially these particle masses are treated as a dynamical quark mass \cite{sss} and further in the deep investigation of QGP, it was replaced by a finite quark mass. This finite value of quark mass depends strongly on temperature \cite{pes2,isrn,kumar}. The thermal dependent quark mass is also advantageous to remove the infrared divergences occurred in the calculations. The earlier work has been modified suitably using the effective quark mass of quasiparticles expressed in Ref. \cite{sri,yy1}. This effective quark mass generated in the suitable environment of relativistic heavy ion collisions produced at RHIC and LHC. The effective quark mass is taken as a linear function of the square value of the current mass, the presence of both terms such as thermal and current mass of quark, and the square value of thermal mass.  
\par 
In this model \cite{pes2,isrn}, a non-interacting system of quasiparticles is described where its existence is based only on temperature. The interaction between quarks and gluons with the neighbouring matter of the medium is responsible for the generation of its mass \cite{ban,bann,bann1}. We incorporated this effective quark mass in our previous analysis of the Equation of State (EoS) \cite{yy1}. Now, we have further adjusted it to account for the influence of a constant magnetic field as well as a time-varying magnetic field, ensuring that all parameters are suitably fitted within the equation. We tried to analyze the equation of state  in  the hot magnetized medium by extending a quasiparticle model in self-consistent manner by fixing zero chemical potential. In a heated and magnetized QGP, three distinct scales emerge, corresponding to the masses of quasiparticles, the temperature, and the strength of the magnetic field. In our approach, we address medium effects by treating quarks and gluons as quasiparticles within the self-consistent model. To accommodate the influence of a robust magnetic field, we extend the model to redefine the effective mass using Landau level quantization for fermions.
\par 	
Firstly, we start with an expression of the effective mass of quasiparticles which is used as \cite{sri,yy1}:
\begin{equation}
{M^2_{eff}}={m_c}^2 + \sqrt{2} {m_c} {m_q}+ {m_q}^2~. 
\end{equation}
In this equation, $m_q$ is taken as mass of the quark dependent on temperature and $m_c$ is the current quark mass. Both values are taken from Ref.~\cite{kumar, poo}.
\par 
The value of finite quark mass is given in equ. ($1$) is defined as \cite{kumar, poo}: 
\begin{equation}
m_{q}^{2}(T) = \gamma_{q} g^{2} T^{2}~.
\end{equation}
We use $\gamma_{q}=1/6$ as used in Ref. \cite{kumar}. The parameter $g^{2}=4 \pi \alpha_{s}~$ is a running coupling constant. 
We have taken care of the parametrization factors incorporated suitably in our calculation to match the results of Lattice QCD. This way, our model is perfectly thermodynamical self-consistent to explain the various features of QGP. The factor $\gamma$ is used as the root mean square value. The main role of this parametrization factor is that it describes the nature of QGP flow.
\par
Now we have modified the effective quark mass in the appropriate environment of strong magnetic field. Since a substantial magnetic field $\sim 10^{15}$ Tesla is generated in the relativistic nuclear collisions of heavy ions at RHIC, LHC, and GSI, it is helpful in investigating the various features of QGP. The recent results help us to see the impact on QGP EoS. The thermodynamic properties of QGP are thus highly influenced by the existence of such large magnetic fields. Thus, the effective quark mass is redefined under the influence of a strong magnetic field produced along z-direction. It is given as \cite{endro1}:

\begin{equation}
M_{eff}^{B}={m_c}^2 + \sqrt{2} {m_c} {m_q}+ {m_q}^2+ eB(2n+s+1)~. 
\end{equation} 
The first three factors on the right hand side are the effective value of quark mass comes from the quasiparticle model. 
\begin{equation}
M_{eff}^{B}={M^2_{eff}}+eB(2n+s+1)~.
\end{equation}
Next, we can consider a constant magnetic field towards the direction of z-axis to describe the magnetic field created in relativistic heavy-ion collisions. 
Due to the relevant degrees of freedom of quarks, these fermion Landau levels could play an interesting role at temperatures above and around the critical temperature. By considering the relativistic Landau levels, the effect of magnetic field can be included. The energy eigenvalues are thus obtained as Landau levels and used under several investigations \cite{bb2}. These studies indicate that there is a dependency of quasiparticle masses on both temperature as well as magnetic field. Finally, we have incorporated the effect of the magnetic field by modifying the effective mass using the relativistic Landau levels. These effective masses are very much significant in computing the thermodynamic observables under the suitable environment of magnetic field.
To this end, due to the quantized motion of charge particles in the environment of magnetic field, the single particle energy spectrum is considered and it is written as \cite{landau}:
\begin{equation}
E=\left[k^{2}+M^2_{eff}+eB(2n+s+1)\right]^{1/2}~,  
\end{equation}

In this context, the notation $n=0,1,2,...$ denotes the principal quantum number associated with permissible Landau levels, while $s=\pm 1$ corresponds to the spin quantum number, with $(+)$ representing quark spin-up and $(-)$ indicating spin-down states. The particle momentum moving along the z-direction of the external magnetic field is denoted by $k$. For simplicity, we adopt the convention $2\nu=2n+s+1$, where $\nu=0,1,2,...$ serves as the quantum number specifying the Landau levels. Even though there is a strong magnetic field created, the thermally generated quarks are feebly jump to the upper Landau levels, so the lowest Landau levels $n=0$ are only populated. We can rewrite the single particle energy eigen value in the form,
\begin{equation}
E=\left[ k^{2}+M^2_{eff}+2\nu eB\right]^{1/2}~, 
\end{equation}
\begin{equation}
E=\left[ k^{2}+M^B_{eff}\right]^{1/2}~. 
\end{equation}
Now it is very easy to convey that the $\nu=0$ state is the only state of singly degenerate while rest other states for $\nu>0$ are doubly degenerate. The energy related to the linear motion through z-axis is quantized, but due to the small energy space, they are treated continuous in nature. Thus, it is used as a single particle energy eigen value. 
\par 
It was studied that not only a strong magnetic field produces in heavy-ion collisions but it also varies slowly with time \cite{t0,t1,yy2}. Authors studied some important properties of this generated magnetic field and it is inferred that the produced magnetic fields stays remaining same during the entire lifetime of plasma evolution. The reason for this is due to the high value of electric conductivity. Later on, it was suggested \cite{t2,t3,t4,t5,t6} that a substantial, in-homogeneous and spatially dispensed magnetic field is certainly produced in collisions of heavy ions in high energy accelerators. The magnitude of the magnetic field is larger at LHC compared to the RHIC. With increasing the proper time, it was observed that at LHC, the strength of the magnetic field decreases more rapidly than that of the RHIC \cite{t7}. According to the quasiparticle model, the particles such as quarks, antiquarks, and gluons would mesh with each other in the existing strong magnetic field environment. While there are various potential mechanisms for the involvement of quarks in the generation of an intrinsic magnetic field, a particularly relevant process occurs when quarks interact with the magnetic field medium. The substantial center-of-mass energy available at RHIC and LHC enables us to examine the effects of a time-varying magnetic field, assumed to be adiabatic in nature \cite{t1}. Referring to equ. ($56$) from Ref. \cite{t1}, the magnetic field depending on time can be expressed as,
\begin{eqnarray}
eB(t)=\hat{y}{\frac{{\alpha}{z}{R}{\sigma}}{t^2}}{\exp(-\frac{{R^2}{\sigma}}{4t})}~,
\end{eqnarray}
where $\sigma$ is referred as the electrical conductivity of QGP, $z$ is taken as the nuclear charge, $R$ is the size of the nucleus and $t$ is time. The presence of a time-varying magnetic field emerges as a crucial factor influencing the Equation of State (EoS) of Quark-Gluon Plasma (QGP). The presence of a time-varying magnetic field emerges as a crucial factor influencing the Equation of State of QGP. This dynamic magnetic field is anticipated to play a pivotal role in shaping the dynamics of QGP, providing insights into the evolution of the system over time and aiding our understanding of its temporal behavior.
 
\section{\bf Equation of state of QGP in the presence of magnetic field}  
The determination of quark gluon plasma equation of state in the existence of a strong magnetic field provides very useful information to the researcher in a diverse fields. A substantial magnetic fields are anticipated to be produced in relativistic heavy ion collisions. At RHIC, the magnetic field strength is contemplated to reach up to $10^{14}$ Tesla, while at the LHC, it surpasses to $10^{15}$ Tesla. This formidable magnetic field intensity opens up new avenues for exploring the behavior of the hot Quark-Gluon Plasma (QGP) system, providing a unique phenomenology. In other astrophysical contexts, neutron stars known as magnetars are predicted to harbor magnetic fields of approximately $10^{10}$ Tesla. This diverse range of magnetic field strengths positions them as valuable factors for the comprehensive study of QGP. Moreover, a magnetic field can interact electro-magnetically or chromo-magnetically which may serve as another useful parameter to study the phase structure of QCD vacuum and phase diagram of QCD on the equal footing with the help of other factors such as chemical potential, thermal dependent quark mass, etc. 
\par
To evaluate the equation of state of QGP, we first compute the total free energy of the system under the influence of a strong magnetic field. The total free energy comprises of three parts: quark (up, down, and strange) free energy, interface free energy term, and gluon free energy. Using total free energy, pressure can be obtained by the negative of the free energy in the thermodynamic limit. So here our main purpose is to calculate the free energy and thereafter pressure in the environment of substantial magnetic field. Now, we define the free energy which is suitably modified for quarks in the existence of magnetic field in Ref. \cite{kumar,yy1}. It is defined as in Ref. \cite{jap}: 
\begin{equation}
F_{q}^{B}=-T g_{q} eB \int \rho_{q} (k) \ln[1+e^{-E /T}] dk~,
\end{equation}
where $\rho_{q}(k)$ is the density of states for quarks \cite{sss}. The degeneracy factor for the quark is taken as $g_{q}$ \cite{yy3}. This equation is a well defined equation for fermions (quarks) where the energy term, $E$, is modified in the presence of an effective quark mass and magnetic field. On the other hand, the gluon free energy part is not affected by the environment of magnetic field. The dominant factor for gluon in a medium is temperature. Thus, there is no change in the gluon free energy term and it is taken as the same as in Ref. \cite{kumar, yy1}.
\begin{figure}[h]
	\centering
	\includegraphics[width=16cm,clip]{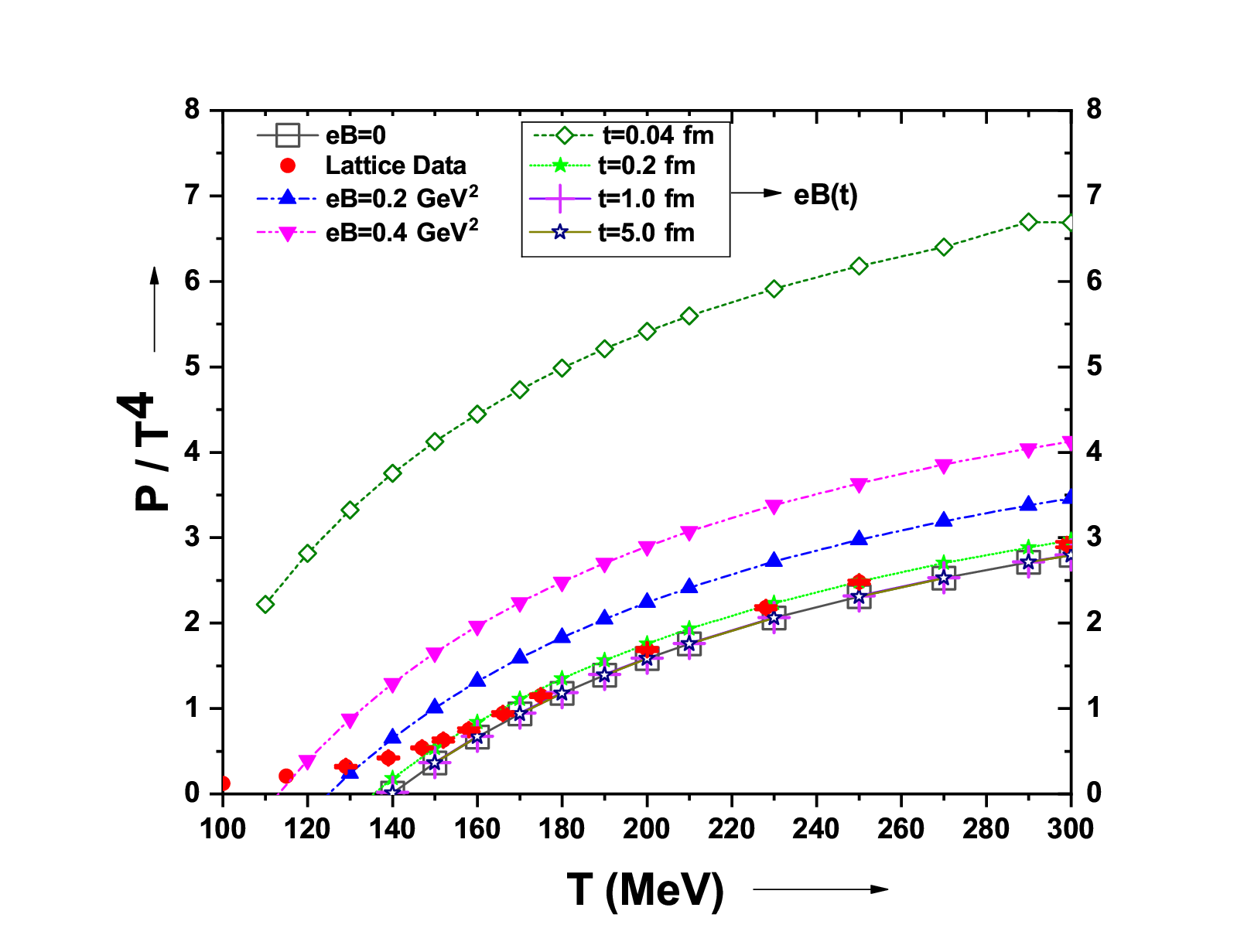}
	\caption{The pressure $(P/T^{4})$ with respect to temperature $(T)$ are shown with time independent and time dependent magnetic field in QGP phase.}
	\label{fig-1}       
\end{figure}
\begin{figure}[h]
	\centering
	\includegraphics[width=16cm,clip]{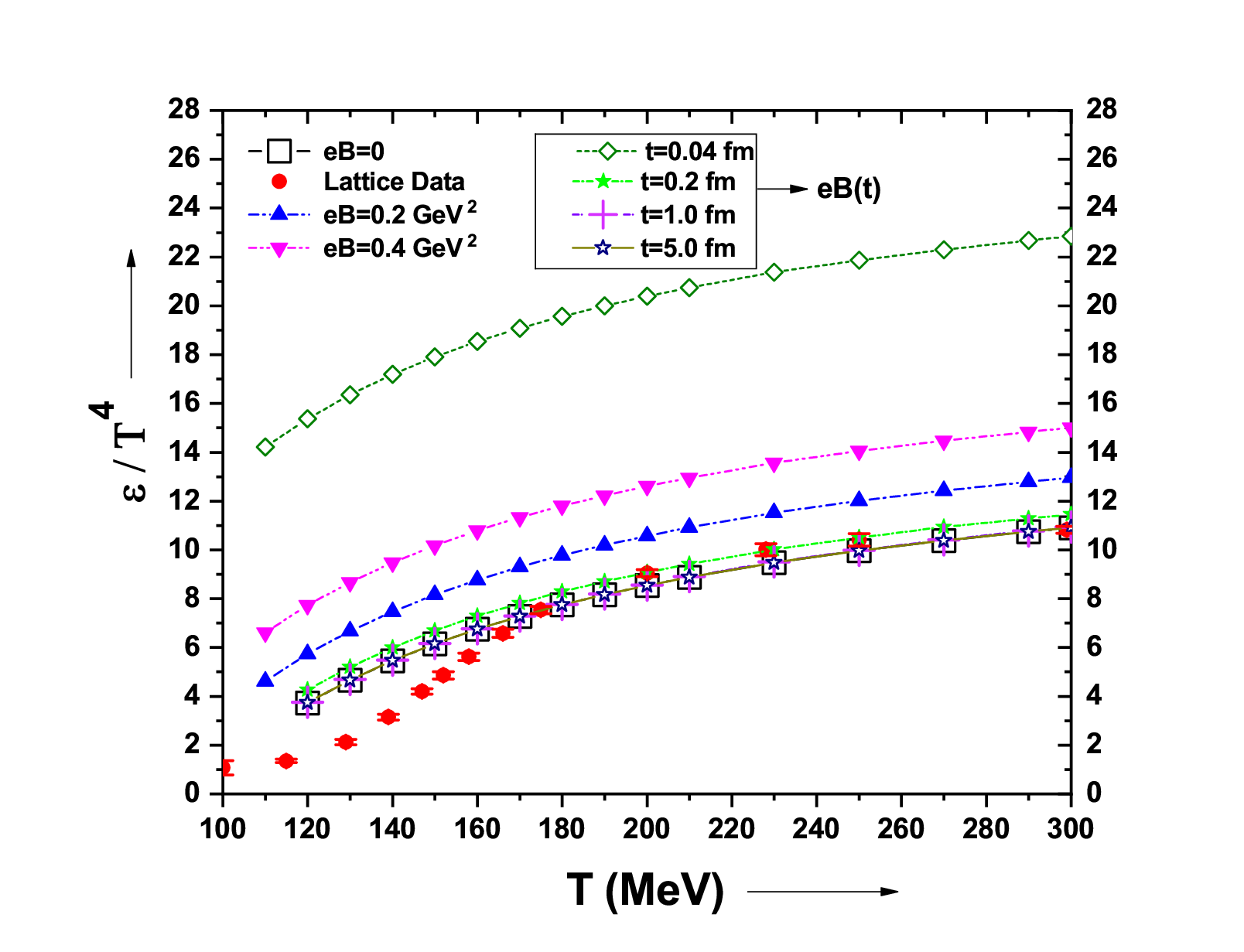}
	\caption{Energy density $(\varepsilon/T^{4})$ with respect to temperature $(T)$ are shown with time independent and time dependent magnetic field in QGP phase.}
	\label{fig-1}       
\end{figure}
\begin{figure}[h]
	\centering
	\includegraphics[width=16cm,clip]{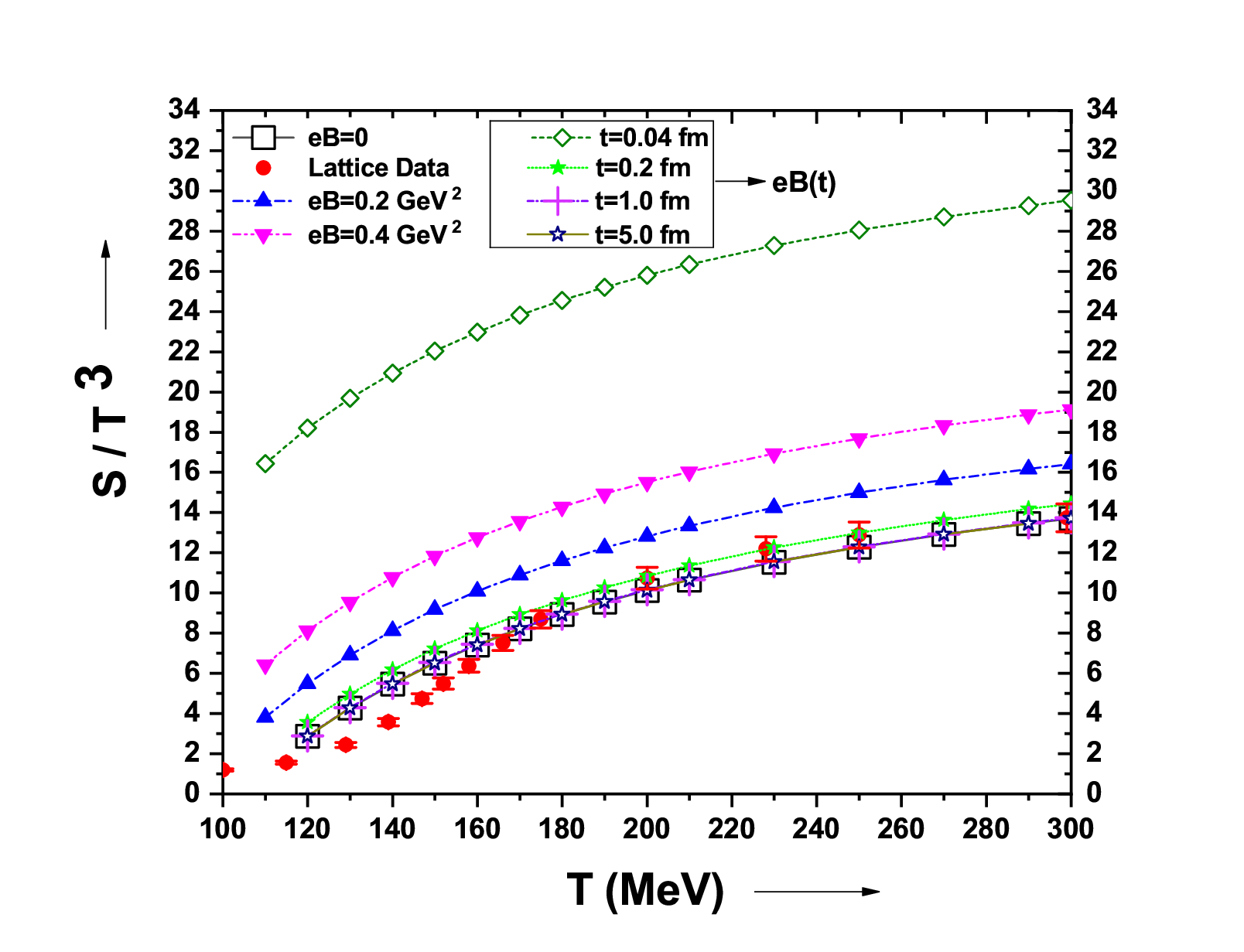}
	\caption{Entropy with respect to temperature $(T)$ are shown with time independent and time dependent magnetic field in QGP phase.}
	\label{fig-1}       
\end{figure}
\begin{figure}[h]
	\centering
	\includegraphics[width=16cm,clip]{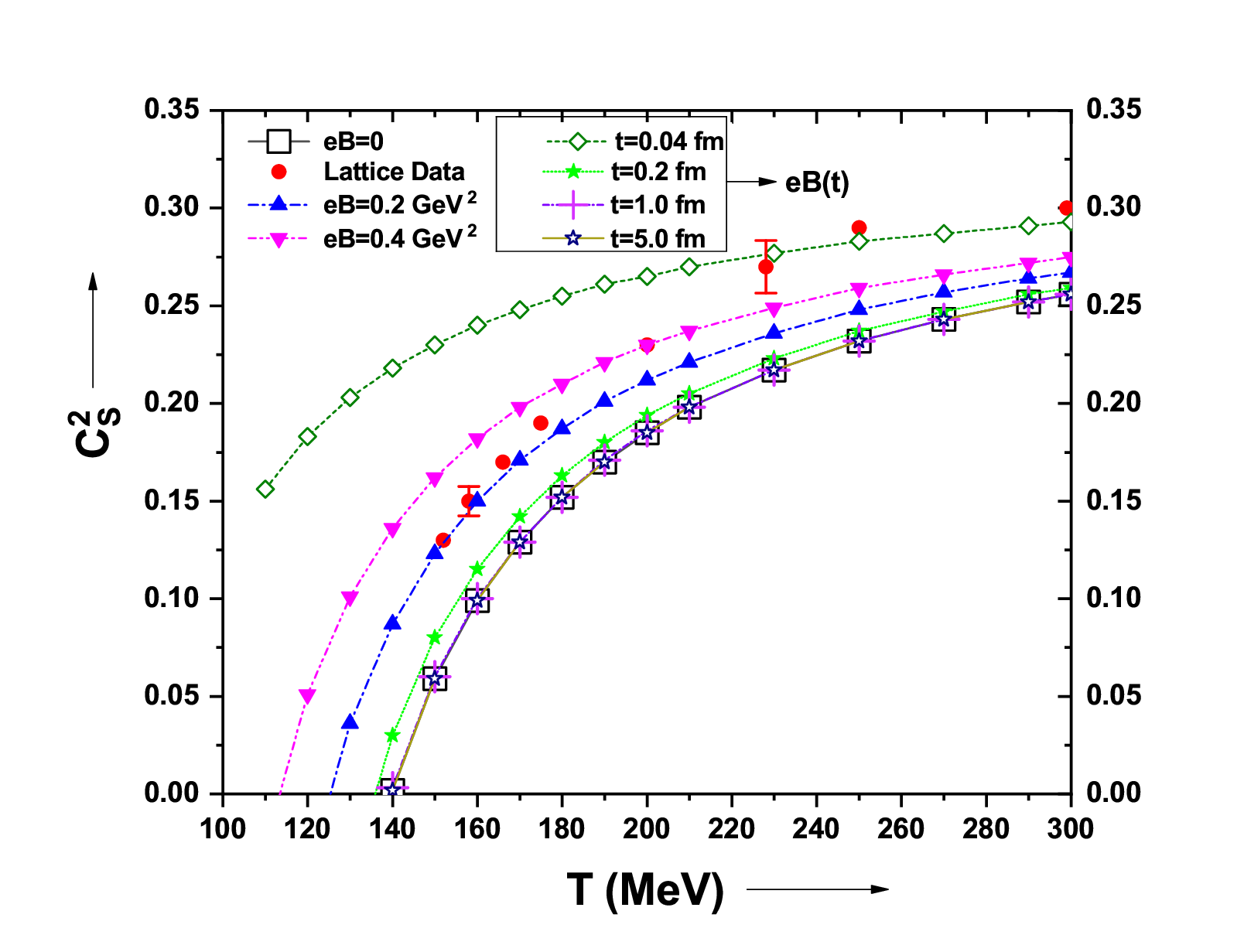}
	\caption{Speed of sound with respect to temperature $(T)$ are shown with time independent and time dependent magnetic field in QGP phase.}
	\label{fig-1}       
\end{figure}
Another important term which separates QGP medium from hadronic medium is the interfacial free energy. It has a significant contribution in replacing the value of bag constant pointed out by Ref. \cite{sss}. 
\par 
Now the total free energy can be evaluated as: 
\begin{equation}
F_{B}^{T}=F_{0}+F_{q}^{B}~.
\end{equation}
Where $F_0$ is the total free energy for quarks, gluon and interface term without magnetic field \cite{kumar,yy1,yy3} and $F_{q}^{B}$ is the free energy for the quark in the occurrence of strong magnetic field \cite{jap}. In the absence of magnetic field, the net value of $F_{B}^{T}$ is retraced as $F_{0}$, i.e., the total free energy in the vanishing magnetic field. The equation tells about the dynamics of QGP flow and helps us to achieve the equation of state. 
Now we can easily calculate the equation of state as the pressure term $P$, energy density term $\varepsilon$, entropy term $S$, and speed of sound term $C_{S}^{2}$ using the value of net free energy $F_{B}^{T}$ in the environment of strong magnetic field. The value of total pressure can suitably obtained using the modified free energy as used in Ref. \cite{jap}:
\begin{equation}
P=-\left( \frac{dF_{B}^{T}}{dv}\right)~.
\end{equation}
Other thermodynamic variables can also be calculated with the support of total pressure. These variables are taken as:
\begin{equation}
\varepsilon=T\frac{dP}{dT}-P~.
\end{equation}
Similarly, the entropy term and the speed of sound term can be calculated as,
\begin{equation}
S=\frac{dP}{dT}~,
\end{equation}
and 
\begin{equation}
C_{S}^{2}= \frac{dP}{d\varepsilon}~.
\end{equation}
\par 
While calculating the thermodynamic observables, temperature and magnetic fields incorporating the effective quark mass are useful parameters to see the impact on equation of state of QGP and we should be very alert in selecting the range of these parameters. In order to see how these thermodynamic variables of a hot matter with three flavors affect the system, we see the variation of pressure value, energy density term, entropy term, and speed of sound term as a function of temperature, with and without magnetic field and with time-varying magnetic field. Free energy evolution provides us the opportunity to calculate all these thermodynamic variables incorporating the effective quark mass. 
We conduct a thorough comparison between our theoretical findings and experimental results. The outcomes derived from the free energy relation enable the computation of various Equation of State (EoS) parameters such as $P/T^{4}$, $\varepsilon/T^{4}$, $S$, and $C_{S}^{2}$. In conclusion, our results bear significance and offer utility in the ongoing exploration of high-energy physics. Thus, in this paper, we use a model to investigate the dynamics of the hot matter of QGP system exposed to the intrinsic and co-existing magnetic field.
\section {\bf Results }
Research in the realms of high-energy heavy-ion collisions, astrophysics, and cosmology has unveiled a common thread: the generation of strong magnetic fields. These magnetic fields manifest in the cores of neutron stars, during the early moments of the universe, and in the heavy ions collisions. In the current study, as we deal with the heavy ion collisions, the time-independent and time-dependent magnetic field are considered for investigating the QGP and to explore its structure via the equation of state of QGP. The giant colliders such as LHC (CERN), RHIC (BNL), FAIR (GSI), etc. are trying to detect QGP formation and its evolution. In recent years, physicists have increasingly acknowledged the ubiquitous presence of a strong magnetic field in various experimental processes, exerting a notable impact on the systems under consideration. Furthermore, observations indicate a substantial modification in the phase structure of Quantum Chromodynamics (QCD) within such environments. Researchers have explored these effects through both theoretical models and Lattice QCD simulations.
\par 
We provide the useful results of EoS of QGP using a simple phenomenological model as the quasiparticle model. We show the plots of thermodynamic variables involving various parameters such as temperature, effective quark mass, time-independent, and time-dependent magnetic field. Thus, the equation of state such as pressure value ($P/T^{4}$), energy density value ($\varepsilon/T^{4}$), entropy term ($S$), and speed of sound value ($C^{2}_{S}$) are shown with respect to temperature ($T$) by varying zero, constant magnetic field and time-varying magnetic field. The explanations of all plots are as follows: 
\par 
Figures $[1]$ and $[2]$ show the pressure term and energy density term with respect to the temperature at zero and constant magnetic field and with time-varying magnetic field incorporating the effective quark mass. In both graphs, we observe that both curve increases exponentially with temperature and approach towards Stefan-Boltzmann (SB) limit for constant magnetic fields as well as for time-varying magnetic fields. Our results are the best matched with the lattice results at zero magnetic field. On comparing the results at fixed non-zero magnetic fields ($0.2 GeV^{2}$, $0.4 GeV^{2}$), it is found that the results of pressure and energy density are much enhanced compared to zero magnetic field. For further investigation, we plot the pressure and energy density with various values of time-dependent magnetic field and compared the results with zero and constant values of magnetic fields. It is interesting to note that for a very short time duration, say, $t=0.04 fm$, the value of pressure and energy density is enhanced very much compared to zero and constant magnetic field and on further increasing the value of time, say, $t=0.2, 1.0, 5.0 fm$, etc. The pressure and energy density curve approach again to the same values of zero magnetic field and matches well with the lattice QCD results. It indicates that in the late time evolution of QGP, the system does not deviate much from its actual phase and tries to retain in the same medium but on the other side, that is, in the initial phase, the system is highly affected and unstable due to the large interaction among the constituent particles. 
\par 
Thus, the involvement of time-dependent magnetic field plays a crucial role and thermodynamic variables are significantly enhanced by considering the early phase of QGP where the time is $\leq$ $0.1 fm$ and on further increasing the time, the system comes back to its same state. It is observed that the magnetic field rapidly decreases in the early time and becomes almost constant in the late time of QGP evolution. So, the results of pressure and energy density for late times are very much similar to the results where the magnetic field is not involved. All results are shown using the effective quark mass which is one of the useful parameters for producing the equation of state of QGP. In view of zero and constant magnetic field, our results are improved as shown in Figures $[1]$ and $[2]$. Results are also much enhanced and large for the initial value of early times, but in the case of late times, it gives the same output as shown for zero magnetic field and for lattice results. The results are similar to our earlier work using the effective quark mass \cite{yy1} at zero magnetic field. Our current results for time-dependent magnetic field are compared with Ref. \cite{jap} at zero and fixed values of the magnetic field.
\par 
Further in Figures $[3]$ and $[4]$, we plot the graph of entropy value and speed of sound value with temperature and with time-independent and time-dependent magnetic fields. In the entropy curve, we found that the outputs at zero magnetic field are lying almost in the same range in lattice results. Thus, the results are almost similar to our earlier work at zero magnetic field using the effective quark mass \cite{yy1}. The results are also verified in the sense that the value of effective quark mass fits very well in the calculation to produce the equation of state of QGP. On the other hand, Lattice QCD results are much enhanced compared to our results for the speed of sound at zero magnetic field. Lattice QCD results for high values of temperature match well with the results of early time of time-varying magnetic field ,i.e., at $t=0.04 fm$. In this case, it is noticed that the thermodynamic variables are affected much in the presence of magnetic field where early times play a key role. So, the equation of state are significantly influenced under time-varying magnetic field, i.e., in the early phase of QGP, while these observables outputs are lying in the same range with the later phase of QGP i.e. $t=0.2, 1.0, 5.0 fm$. Entropy observable also does not affect with large value of time-varying magnetic field and hence shows similar results as for zero magnetic field. Among all observables, the speed of sound for large values of temperature is the only observable which matches well with the results of lattice QCD simulation at the early phase of QGP, i.e., t $\leq$ $0.1 fm$.
\par 
Ultimately, our model's outcomes remain consistent with our prior findings concerning the effective quark mass in scenarios with zero magnetic field and significantly large time-varying magnetic fields. Conversely, the EoS of QGP exhibits a substantial influence from constant magnetic fields and lower values of time-varying magnetic fields. Notably, predictions from other models regarding the EoS of QGP align with earlier results obtained from Lattice QCD \cite{brau, sri, ban, bann, jap}.
 
\section{\bf Conclusion}
\par  
In this article, we work on thermodynamic observables of a hot QGP system which plays an important role in the presence of a constant and time-dependent strong magnetic field produced in the collisions of ultra-relativistic heavy ions. Since these observables have contributions from constituent particles such as quarks, they are notably affected by the strong magnetic field while the gluon part does not affect in the environment of strong magnetic field and it depends only on temperature. In this work, we obtain the important QGP equation of state such as pressure term, energy density term, entropy term, and the speed of sound term. The time-independent and time-dependent magnetic fields show significant output and thus the results are compared with zero magnetic field and with Lattice QCD results. The results are also compared with other works. 
\par 
The current observations may be useful implications on the hot system of QGP and the expansion dynamics of QGP medium at RHIC, LHC, and GSI show interesting features in the presence of magnetic field. These results could have useful outcomes for the study of various signatures of QGP. The results are shown for a variety of thermodynamic observables which indicate that the equation of states are significantly affected not only with the constant values of magnetic field but also useful with time-varying magnetic field especially in the early phase of QGP formation. Our results are very much similar for the late phase time of QGP under the effect of magnetic field. Thus, for late time, the results of EoS match exactly with the results of zero magnetic field and similar to the lattice QCD results except the results of speed of sound. The speed of sound enhances significantly during the early time and matches well with the Lattice QCD data. 
\par 
In conclusion, our model, incorporating the effective quark mass under the influence of both time-dependent and time-independent magnetic fields, yields improved results compared to those obtained in the absence of a magnetic field and aligns well with the latest EoS findings. This enhancement is particularly evident in the exploration of the phase structure of QGP. The model proves versatile in investigating various properties of QGP under strong magnetic fields. It is imperative to underscore the significance of time-varying magnetic fields, especially during the early phases of QGP, where EoS results are markedly affected. This presents an avenue for further exploration by researchers. Our results stand out compared to alternative theoretical approaches, offering advantages in the measurement of QGP equation of state. The consequential implications of these findings, both with and without time-varying magnetic fields, have broad phenomenological relevance across various domains of high-energy physics.
 
\subsection{Acknowledgments}
Authors would like to thank Principal, Hansraj College, University of Delhi for providing the necessary facilities for pursuing research in the college.
\\
\\
\newline 
\textbf{Funding Statement}
\newline
The author declare that the work is performed as part of the employment of the author, hansraj college, university of delhi and they have no known competing financial interests or personal relationships that could have appeared to influence the work reported in this paper.
\\
\newline 
\textbf{Contributions}
\newline 
All authors have equally contributed and accepted responsibility for this submitted manuscript’s entire content and approved submission.
\newline 
\\
\textbf{Data Availability}
\newline 
The original contributions presented in the study are included in the article/supplementary material, and further inquiries can be directed to the corresponding authors.
\newline 
\textbf{Conflict of Interests}
\newline 
The authors declare that they have no conflicts of interest regarding the publication of this paper.

\newpage 
\par

\end{document}